\documentclass[twocolumn]{aastex62}

\usepackage{amsmath}
\usepackage{xcolor}
\usepackage{mathrsfs}
\usepackage[caption=false]{subfig}
\usepackage{float}
\usepackage{multirow}
\usepackage{enumitem}

\shorttitle{Black hole mass function and its redshift evolution by ET}
\shortauthors{Ding et al.}

\begin{document}

\newcommand{\mbh}{$\mathcal M_{\rm BH}$}
\newcommand{\cmass}{${\cal M}_0$}
\newcommand{\dl}{$d_L$}
\newcommand{\mone}{$m_1$}
\newcommand{\mtwo}{$m_2$}
\newcommand{\snr}{$\rho$}

\title{Black hole mass function and its evolution -- the first prediction for the Einstein Telescope}

\correspondingauthor{Zong-Hong Zhu}
\email{zhuzh@whu.edu.cn}

\author
{Xuheng Ding}
\affiliation{School of Physics and Technology, Wuhan University, Wuhan 430072, China}
\affiliation{Department of Physics and Astronomy, University of California, Los Angeles, CA, 90095-1547, USA}


\author
{Kai Liao}
\affiliation{School of Science, Wuhan University of Technology, Wuhan 430070, China}

\author
{Marek Biesiada}
\affiliation{Department of Astronomy, Beijing Normal University, Beijing 100875, China}
\affiliation{National Centre for Nuclear Research, Pasteura 7, 02-093 Warsaw, Poland}

\author
{Zong-Hong Zhu}
\affiliation{School of Physics and Technology, Wuhan University, Wuhan 430072, China}
\affiliation{Department of Astronomy, Beijing Normal University, Beijing 100875, China}

\begin{abstract}
The knowledge about the  black hole mass function (BHMF) and its evolution would help to understand the origin of the BHs and how BH binaries formed at different stages of the history of the Universe. We demonstrate the ability of future third generation gravitational wave (GW) detector -- the Einstein Telescope (ET) to infer the slope of the BHMF and its evolution with redshift.  We perform the Monte Carlo simulation of the measurements of chirp signals from binary BH systems (BBH) that could be detected by ET, including the BH masses and their luminosity distances (\dl). We use the mass of a primary black hole in each binary system to infer the BHMF as a power-law function with slope parameter as $\alpha$. Taking into account the bias that could be introduced by the uncertainty of measurements and by the selection effect, we carried out the numerical tests and find that only one thousand of GW events registered by ET ($\sim1\%$ amount of its yearly detection rate) could accurately infer the $\alpha$ with a precision of $\alpha\sim0.1$. Furthermore, we investigate the validity of our method to recover a scenario where $\alpha$ evolves with redshift as $\alpha(z) = \alpha_0 + \alpha_1\frac{z}{1+z}$. Taking a thousand of GW events and using \dl\ as the redshift estimator, our tests show that one could infer the value of evolving parameter $\alpha_1$ accurately with the uncertainty level of $\sim0.5$. Our numerical tests verify the reliability of our method. The uncertainty levels of the inferred parameters can be trusted directly for the several sets of the parameter we assumed, yet shouldn't be treated as a universal level for the general case.
\end{abstract}


\section{Introduction} \label{sec_intro}

The masses of astrophysical black holes (BHs) are known to cover a wide range from stellar-mass to supermassive level ($\sim10^{10} M_{\odot}$). The discovery of
coalescing binary black holes (BBHs) in LIGO gravitational-wave (GW) detectors is a substantial evidence of stellar-mass BHs \citep{Abbott2016}, while supermassive BHs are supposed to exist in the centers of almost all the galaxies \citep{Lynden-Bell1969, Kormendy1995}.
GWs provide a direct way to study the inspiralling BBH systems, enabling one to derive their basic parameters including mass, spin and luminosity distances~\citep{Abbott2017phy, Abbott2018}. This creates the opportunity not only to measure the properties of BHs ~\citep{Abbott2018b}, but also answer some fundamental questions concerning cosmography~\citep{Liao2017, Ding2019, Cai2017}, the GW speed~\citep{Fan2017, Collett2017} or the strong lensing of GWs ~\citep{Ola2013, Biesiada2014, Ding2015}. 

Nevertheless, it is still unclear of how the BHs are formed \citep{Fryer1999, Fryer2001, Mirabel2016}. In particular, the number and mass distribution of stellar-mass BHs in the Universe still need to be clarified.
The recent detections of GW events have brought us a new era of gravitational wave astronomy \citep[e.g.,][]{Abbott2016, Abbott2016_sum, Abbott2018} and opened up a  brand new possibility concerning studying BBH system formation channels.
At present, however, observations cannot firmly select the basic formation scenarios like the evolution of isolated pairs of stars \citep{Bethe1998, Portegies1998}, chemically homogeneous evolution \citep{Marchant2016, deMink2016}, dynamic binary formation in dense clusters \citep{Portegies2000, Kulkarni1993} and other channels introduced in \citet{Abbott2018b}.
The inference of the distributions of BH mass could be the key to distinguish these scenarios and help to address questions including the physical process and evolutionary environment of binary BH formation.

Current GWTC-1 catalog of binary coalescences detected by LIGO/Virgo GW interferometers includes ten BH-BH binaries and one NS-NS (GW170817) binary \citep{Abbott2018}.  Assuming the BH mass function (BHMF) parametrized as a two-sided truncated power-law, \citet{Kovetz2017PhRvD} estimated that further LIGO measurements would provide thousands of BBHs and constrain the BHMF slope parameter $\alpha$ at 10\% precision. More recently, the LIGO collaboration has used ten BBH merger events and constrained the BHMF power-law index to $\alpha~=~1.6\substack{+1.5\\-1.7}$ (90\% credibility) \citep{Abbott2018b}.
In the next decade, the number of detected coalescences of BBH systems is expected to be increasing rapidly with the improvements of the detector sensitivities. Especially, the third-generation gravitational wave detector Einstein Telescope (ET) is capable of detecting $10^4-10^8$ coalescing BBHs per year \citep{Abernathy2011}. Moreover, since this instrument would detect the GW events from the distant Universe up to $z\sim17$ \citep{Abernathy2011}, the wide redshift range of the BBH inspiral events enable us to study the $\alpha$ as a function of redshift. In this study, we use the Monte Carlo (MC) approach to simulate the GW events from BBH mergers that could be measured by the ET. We construct a mock BBH merger catalog to examine their ability to constrain the BHMFs, taking into account the data noise level and selection bias realistically.

This paper is organized as follows. In Section~\ref{sec_simulation} we describe the simulation of the BBH inspiral events detectable by ET using the Monte Carlo approach. In this section, we assume the initial assumptions for the BH mass function used further as true values to be recovered from the data. In Section~\ref{sec_theory}, we introduce the theoretical framework to reconstruct the BHMFs, considering the noise realization and the selection effects. Furthermore, we make a further step by considering the power-law index $\alpha$ as a function of redshift and explore the way to use luminosity distance as redshift estimator and detect such evolution. We present our results in the Section~\ref{sec_result}. The discussion and conclusions are given in the Section~\ref{sec_summary}. Throughout this paper, we assume a standard concordance cosmology with $H_0= 70$ km s$^{-1}$ Mpc$^{-1}$, $\Omega{_m} = 0.30$, and $\Omega{_\Lambda} = 0.70$.

\vspace{1cm}
\section{Data simulation} \label{sec_simulation}
In this section we describe the simulation of a realistic mock catalog of GW signals from BBHs detectable by future ET interferometric detector. Numerical predictions of BBH inspirals detectable by ET
have been discussed in many works, and it has been forecasted that the yearly detection rate of BBHs would be of order $\sim10^{4-8}$ \citep{Abernathy2011} or at least $\sim10^{5}$ according to less optimistic yet realistic scenarios \citep{Ola2013, Biesiada2014}. More recently, \citet{Yang2019} developed the approach of a Monte Carlo (MC) simulation to predict the detection rate by explicitly considering each BBH inspiral event sampled from the outcome of the population synthesis model, which provides the way to mimic a realistic BBH GW catalog. The backbone of this approach is to use random seeds to build up a mock universe which includes a sufficient volume of BBH inspiral events with essential parameters that related to this study. We refer the readers for the details in \citet[][Section 2, therein]{Yang2019}  and briefly recall the key points below.

\subsection{Detection Criteria} \label{subsec_criteria}
For a specific BBH inspiral event at redshift $z_s$, the ET's corresponding signal-to-noise ratio $\rho$ is defined as \citep{Abernathy2011}:

\begin{equation} \label{SNR}
\rho = 8 \Theta \frac{r_0}{d_L(z_s)} \left( \frac{(1+z){\cal M}_0}{1.2 M_{\odot}} \right)^{5/6}
\sqrt {\zeta(f_{max})},
\end{equation}
where $r_0$ is the detector's characteristic distance parameter and $\zeta(f_{max})$ is the dimensionless function reflecting the overlap between the GW signal and the ET's effective bandwidth. For simplicity, we followed  \citet{Taylor2012} and approximated $\zeta(f_{max})$ as unity. ${\cal M}_0$ is the intrinsic chirp mass defined as $ {\cal M}_0 = \frac{(m_1m_2)^{3/5}}{(m_1+m_2)^{1/5}}$, where \mone\ and \mtwo\ are the
respective masses of the BBH components. $\Theta$ is the orientation factor determined by four angles as \citep{Finn93}:
 \begin{equation} \label{Theta}
 \Theta = 2 [ F_{+}^2(1 + \cos^2{\iota} )^2 + 4 F_{\times}^2 \cos^2{\iota} ]^{1/2},
 \end{equation}
where: $F_{+} = \frac{1}{2} (1 + \cos^2{\theta}) \cos{2\phi} \cos{2 \psi} - \cos{\theta} \sin{2 \phi} \sin{ 2 \psi}$, and
$F_{\times} = \frac{1}{2} (1 + \cos^2{\theta}) \sin{2\phi} \cos{2 \psi} + \cos{\theta} \sin{2 \phi} \cos{ 2 \psi}$ are so-called antenna patterns. The four angles ($\theta, \phi, \psi, \iota$) describe respectively the
direction to the BBH system relative to the detector and the binary orientation relative to the line of sight between it and the detector.
They are independent and one can assume that $(\cos\theta, \phi/\pi, \psi/\pi, \cos\iota)$ distributed uniformly over the range $[-1, 1]$. The GW signal is considered as detectable if its \snr\ is over the detecting threshold, i.e., $\rho > \rho_0 = 8$.

\subsection{Monte Carlo Approach} \label{MC}

We aim to build up a sufficient volume of BBH systems in the mock universe by randomly generating the key parameters for each BBH system as specified below. First key parameter is the redshift $z_s$.  We sample the merging BBH systems according to the yearly merger rate in a redshift interval  $[z_{s}, z_{s}+dz_{s}]$:
 \begin{equation}
 d\dot{N} (z_s)=4\pi\left(\frac{c}{H_{0}}\right)^3\frac{\dot{n}_{0}(z_{s})}{1+z_{s}}\frac{\tilde{r}^2(z_{s})}{E(z_{s})}dz_{s}.
 \end{equation}
where the intrinsic BBH merger rate $\dot{n}_{0}(z_{s})$ is the one predicted by the population synthesis model (using {\tt StarTrack} code\footnote{The data is taken from the website \url{http://www.syntheticuniverse.org}.}) in \citet{Dominik13}, $\tilde{r}(z_{s})$ is the dimensionless comoving distance to the source, and $E (z_s)$ is
the dimensionless expansion rate of the universe at redshift $z_s$.
Other key parameters include the four angles $(\theta, \phi, \psi, \iota)$ in the Equation~(\ref{Theta})  and the masses of each BH in the binary system (i.e., \mone\ and \mtwo).
For the purpose of randomly generating the BH masses, we follow the previous works \citep{Kovetz2017PhRvD, Abbott2018b, Fishbach2018} and assume that \mone\ follows a power-law distribution with a hard cut at both maximum and minimum mass:
 \begin{equation} \label{equ_powlaw}
P(m_1|\alpha, M_{max}, M_{min}) = m_1^{\alpha} \mathcal{H}(m_1-M_{min}) \mathcal{H}(M_{max}-m_1),
 \end{equation}
where $\mathcal{H}$ is the Heaviside step function. Then, the secondary mass, \mtwo\, is 
sampled from a uniform distribution
between $[M_{min}, m_1]$. Let us note that, we only take the \mone\ to reconstruct the BHMF, thus the assumption of the distribution for \mtwo\ actually does not affect the inference for the shape of BHMF.
For the purpose of the simulation, however, all these parameters are necessary to determine the value of $\Theta$ and ${\cal M}_0$ in Equation~(\ref{SNR}). We combine them with their redshift $z_s$ to generate the $\rho$ of each BBH inspiraling system. We only collect the events which have $\rho > \rho_0 = 8$, meaning that those events with $\rho < 8$ are too faint to detect.

Concerning the BHMF we consider two scenarios. In the first scenario, the exponent $\alpha$ is constant, hence the shape of the BHMF is fixed throughout the redshfit range probed by the ET. In the second scenario, we consider that $\alpha$ varies as a function of redshift according to:
 \begin{equation} \label{equ_alphaz}
\alpha(z) = \alpha_0 + \alpha_1\frac{z}{1+z} ,
 \end{equation}
so that the $\alpha(z)$ would transform gradually from $\alpha_0$ to $\alpha_0+\alpha_1$ through low$-z$ to high$-z$. We do not have any clear physical guidance of how could $\alpha$ evolve  with redshift and in particular which analytical expression would describe it reliably. Therefore the above ansatz is actually a suitable form of the first order Taylor expansion of $\alpha$ as a function of the scale factor $a$ (around the present value $a(t_0)=1$, where $a(t) = \frac{1}{1+z}$).

\subsection{Estimation of Parameter Error} \label{sec_noiselevel}
We aim to produce the mock dataset of the future GW events representative of the ET  measurements. In order to consider the measurement uncertainties in the realistic way, we distribute random statistical uncertainties into the simulated data as described below.

The quantities measurable from the BBH inspiral waveform comprised of \dl, redshifted chirp mass $(1+z){\cal M}_0 $ and $\rho$. Individual masses \mone\ and \mtwo\ are derived from the combination of the chirp mass and the total mass \mone+\mtwo, which can also be extracted from the chirp waveform.
Let us note that physical quantities inferred from the LIGO detections had asymmetric upper and lower uncertainty limits, hence they followed the  ${\it skewed}$  distributions. Therefore, instead of the symmetric Gaussian distribution, we assume that the simulated mock measurements follow the Log-Normal distribution with the standard deviation of  ${\cal M}_0$, \dl, and \mone\ equal to 0.17, 0.35 and 0.2, respectively.
For instance, if the $m_{1,fid}$ is the true value for \mone, the probability density used to simulate the measured value is:
 \begin{equation} \label{equ_lognorm}
P(m_1) = \frac{1}{m_1 \sigma_{m_1} \sqrt{2\pi}} exp \left[- \frac{log(m_1)-log(m_{1,fid})}{2\sigma_{m_1}} \right].
 \end{equation}
We set up the uncertainties for ${\cal M}_0$, \dl, and \mone\ by taking results of \citet{Ghosh2016} as the reference, who explored the expected statistical uncertainties with which the parameters of black hole binaries can be measured from GW observations by next generation ground-based GW observatories. Note that the assumed uncertainty level of these quantities only affect the uncertainty of the inferred parameters (i.e. the precision) and wouldn't affect the validity test of our method (i.e. the  accuracy).

Having clarified the MC approach and defined data the uncertainty level, we are capable of producing the mock GW dataset. For demonstrating propose, we list an example of one thousand BBH inspiral events as simulated in one realization of the MCMC seeding process. 

\begin{deluxetable}{lcccc}
\tablecolumns{5}
\tabletypesize{\footnotesize}
\tablewidth{0pt}
\tablecaption{Illustration of the mock GW catalog}
\tablehead{
\colhead{Object ID} &
\colhead{\mone}&
\colhead{Luminosity Distance} &
\colhead{Chirp Mass} &
\colhead{SNR}
\\
\colhead{} &
\colhead{($M_\odot$)}&
\colhead{(Mpc)} &
\colhead{($M_\odot$)} &
\colhead{(\snr)}
\\
\colhead{(1)} &
\colhead{(2)} &
\colhead{(3)} &
\colhead{(4)} &
\colhead{(5)}
}
\startdata
ID1 & $8.14\substack{+1.80\\-1.48}$  & $25356.8\substack{+5614.1\\-4596.4}$  & $9.91\substack{+2.19\\-1.80}$ & 18.512 \\
ID2 & $43.93\substack{+9.73\\-7.96}$  & $12956.9\substack{+2868.7\\-2348.7}$  & $28.64\substack{+6.34\\-5.19}$ & 46.953 \\
ID3 & $5.05\substack{+1.12\\-0.92}$  & $8617.6\substack{+1908.0\\-1562.1}$  & $3.60\substack{+0.80\\-0.65}$ & 9.836 \\
ID4 & $13.93\substack{+3.08\\-2.53}$  & $47473.1\substack{+10510.7\\-8605.4}$  & $10.97\substack{+2.43\\-1.99}$ & 17.563 \\
ID5 & $11.37\substack{+2.52\\-2.06}$  & $21485.0\substack{+4756.8\\-3894.6}$  & $10.68\substack{+2.36\\-1.94}$ & 31.107 \\
ID6 & $21.67\substack{+4.80\\-3.93}$  & $9446.1\substack{+2091.4\\-1712.3}$  & $7.64\substack{+1.69\\-1.39}$ & 31.624 \\
ID7 & $32.20\substack{+7.13\\-5.84}$  & $69317.4\substack{+15347.1\\-12565.1}$  & $19.35\substack{+4.28\\-3.51}$ & 13.496 \\
ID8 & $7.73\substack{+1.71\\-1.40}$  & $25104.5\substack{+5558.2\\-4550.7}$  & $6.68\substack{+1.48\\-1.21}$ & 9.612 \\
ID9 & $35.63\substack{+7.89\\-6.46}$  & $16321.8\substack{+3613.7\\-2958.6}$  & $10.96\substack{+2.43\\-1.99}$ & 16.202 \\
ID10 & $18.78\substack{+4.16\\-3.40}$  & $54375.1\substack{+12038.8\\-9856.5}$  & $15.16\substack{+3.36\\-2.75}$ & 29.232 \\
... & $...$ & ... & ... & ...\\
ID991 & $29.40\substack{+6.51\\-5.33}$  & $40765.1\substack{+9025.5\\-7389.5}$  & $29.79\substack{+6.59\\-5.40}$ & 28.134 \\
ID992 & $6.41\substack{+1.42\\-1.16}$  & $16026.7\substack{+3548.4\\-2905.2}$  & $4.54\substack{+1.01\\-0.82}$ & 17.550 \\
ID993 & $10.27\substack{+2.27\\-1.86}$  & $33207.9\substack{+7352.3\\-6019.6}$  & $8.57\substack{+1.90\\-1.55}$ & 8.037 \\
ID994 & $21.26\substack{+4.71\\-3.85}$  & $56088.6\substack{+12418.2\\-10167.1}$  & $16.01\substack{+3.54\\-2.90}$ & 19.871 \\
ID995 & $14.35\substack{+3.18\\-2.60}$  & $29022.0\substack{+6425.6\\-5260.8}$  & $13.50\substack{+2.99\\-2.45}$ & 14.230 \\
ID996 & $22.50\substack{+4.98\\-4.08}$  & $49038.7\substack{+10857.3\\-8889.2}$  & $18.48\substack{+4.09\\-3.35}$ & 11.636 \\
ID997 & $5.42\substack{+1.20\\-0.98}$  & $23537.5\substack{+5211.3\\-4266.6}$  & $4.54\substack{+1.01\\-0.82}$ & 12.562 \\
ID998 & $25.07\substack{+5.55\\-4.55}$  & $19555.4\substack{+4329.6\\-3544.8}$  & $18.81\substack{+4.16\\-3.41}$ & 17.846 \\
ID999 & $5.70\substack{+1.26\\-1.03}$  & $24024.9\substack{+5319.2\\-4355.0}$  & $3.61\substack{+0.80\\-0.65}$ & 23.020 \\
ID1000 & $39.23\substack{+8.69\\-7.11}$  & $31835.1\substack{+7048.4\\-5770.7}$  & $22.31\substack{+4.94\\-4.04}$ & 15.884 \\
\enddata
\label{tab_GW_mock_data}
\tablecomments{The catalog of simulated thousand BBH inspiral events is used to test the inference of the BHMF from the data attainable in forthcoming next generation GW detector -- the ET. The reported values are the medians, with errors corresponding to the 16th and 84th percentiles, assuming $\alpha=1.6$, $M_{min}~=~5M_{\odot}$, $M_{max}~=~50M_{\odot}$. Note that this mock data is re-simulated every time in each realization}.
\end{deluxetable}

\vspace{1cm}
\section{Theoretical Framework}  \label{sec_theory}
In this section, we describe the fitting procedure for the parameterized BHMFs.
In principle, the modeling for a dataset which follows a power-law distribution as Equation~(\ref{equ_powlaw}) is very straightforward. To derive the posterior of the parameters, one only needs to combine all the measured median values together in a joint likelihood:
 \begin{equation} \label{equ_lik_powlaw}
 P(\alpha, M_{max}, M_{min}|m_{1}) \propto  \prod_{i=1}^{total} P(m_{1,i}|\alpha, M_{max}, M_{min})
 \end{equation}
 where $m_{1,i}$ is the primary mass inferred from the $i$~-~$th$ GW event.
However, the median values of simulated \mone, as shown in Table~\ref{tab_GW_mock_data}, actually deviate from the initial power-law distribution. This deviation stems from several effects that exist in reality. In Section~\ref{sec_likelihood_noise} and \ref{sec_likelihood_sf}, we introduce them and explore the ways to account for them.

\subsection{Measurement Uncertainty}\label{sec_likelihood_noise}
\begin{figure}
\includegraphics[width=1.05\linewidth]{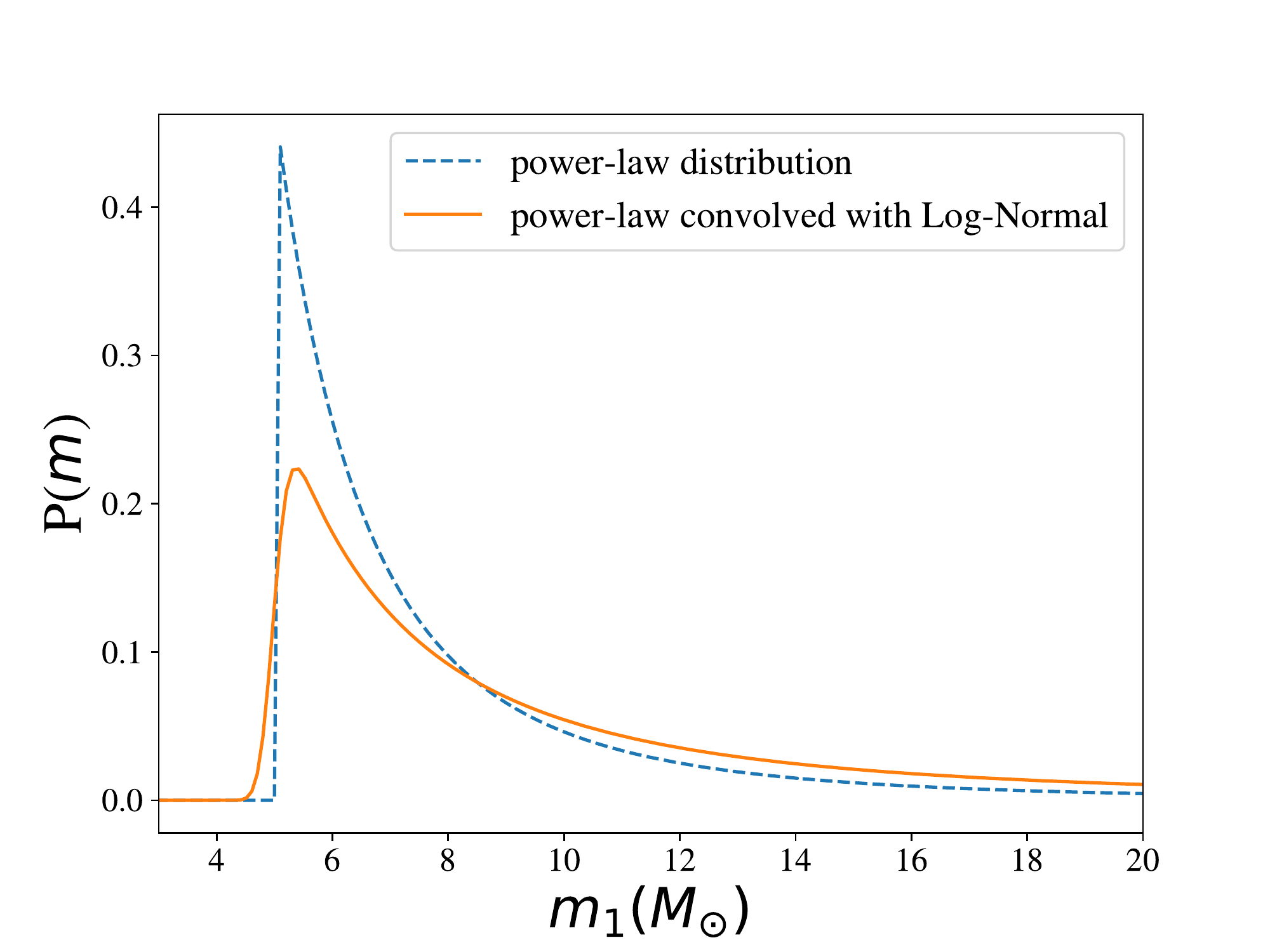}
\caption{
Figure illustrating the convolution of a power-law distribution with a Log-Normal distribution having $\sigma = 0.2$. One can see that the convolution make distribution shallower, smoothes the breaking edge at $m_1 = 5 M_{\odot}$ and makes the slope less steep.
}
\label{fig:result_slope}
\end{figure}

The intrinsic value of primary BH mass (i.e., $m_{1,fid}$) follows a power-law distribution, however the measured  \mone\ is scattered by the Log-Normal distribution which does not follow a power-law function anymore \citep{Koen2009}. In theory, if the event $X$ follows a power-law distribution and its observed values are subject to the  Log-Normal uncertainty, then the observed event $X + e$,   with $e$ denoting the error (uncertainty), is distributed according to the convolution of the power-law and Log-Normal distribution. Assuming that the noised data follow the Log-Normal distribution, we convolved the intrinsic power-law to describe likelihood as:
 \begin{equation} \label{equ_lik_conv}
 P(\alpha, M_{max}, M_{min}|m_{1}) \propto  \prod_{i=1}^{total} \hat{P}(m_{1,i}|\alpha, M_{max}, M_{min}),
 \end{equation}
where the $\hat{P}$ is the a power-law function convolved with the Log-Normal distribution using the standard deviation as 0.2 as we assumed. We illustrate the effect of such convolution in the Figure~\ref{fig:result_slope}.

\subsection{Selection Effect}\label{sec_likelihood_sf}
The GW observations have a tendency to discover more significant events, known as Malmquist bias. For example, the GW systems with higher values \mone\ tend to produce stronger signals and thus have a higher probability to be detected. As a result, the final BHMFs would be biased to the high mass end, if this effect is not correctly taken into account.

\begin{figure}
\includegraphics[width=1.0\linewidth]{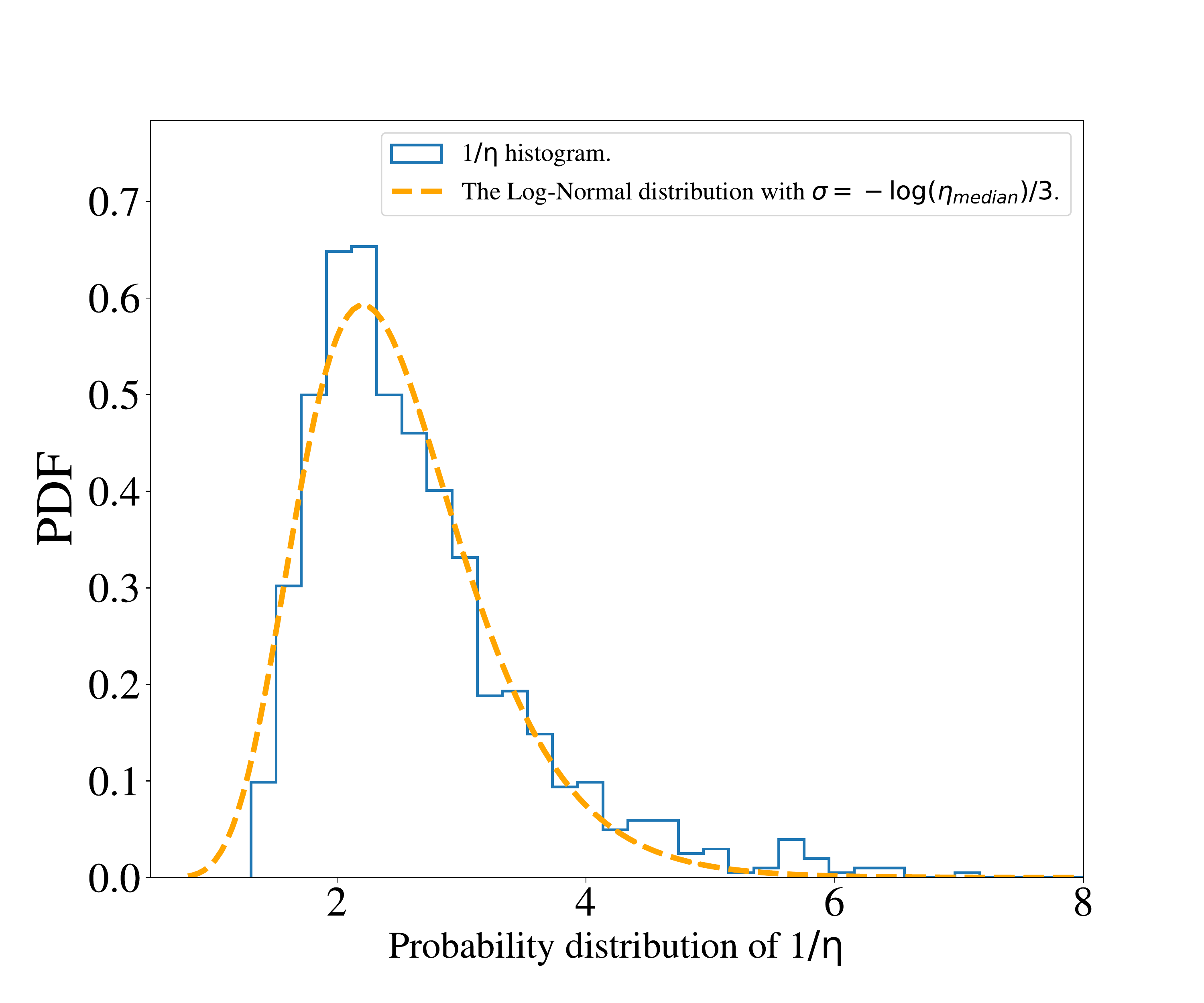}
\caption{
Assuming a set of \dl\ and \cmass\ following the Log-Normal distribution, we randomly produce the corresponding histogram of the $1/\eta$ in order to assess its probability distribution. The result shows that the skewed distribution could be well described by a Log-Normal distribution with $\sigma = -\log(\eta_{\rm median})/3$.
}
\label{fig_eta}
\end{figure}

To overcome this selection effect, we introduce the selection factor $\eta$ for the GW event, which is the detecting probability of one event in a repeated simulation.
The meaning of this factor $\eta$ is straightforward -- if one GW event has $\eta=0.2$, it means that this event has 80\% probability of being be missed. In other words, four equivalent events would have been missed. Thus, for this event, one needs to re-calibrate this influence by enhancing the likelihood by a power of 5 (i.e., $L^{1/0.2} = L^5$) to recover the intrinsic probability value. Hence, to account for the selection effect, we calculate the likelihood as:
 \begin{equation} \label{equ_lik_sf}
 P(\alpha, M_{max}, M_{min}|m_{1}) \propto  \prod_{i=1}^{total} \hat{P}(m_{1,i}|\alpha, M_{max}, M_{min})^{1/\eta},
 \end{equation}
where $\eta$ is directly determined by the probability distribution of $\rho$, i.e. $\eta = P(\rho>8)$. In order to use the Equation~(\ref{SNR}) to calculate $\rho$, the distribution function of $\Theta$ is taken from the MC simulations; the \cmass\ and \dl\ are adopted from the mock dataset as demonstrated in Table~\ref{tab_GW_mock_data}. Yet, the redshift $z_s$ is the unknown parameter since it is non-measurable in the GW detectors; one can only take the \dl\ as redshift estimator.
Note that the observed \dl\ and \cmass\ are both considered to have random noise which follows the asymmetric distribution (i.e., Log-Normal). Thus, the intrinsic probability distribution (not the errors to be convolved) of their product, and thus of $\eta$ is also asymmetric. Considering the random distributions of the \dl\ and \cmass, we performed the numerical tests and found that the distribution of $1/\eta$ could be well described by the Log-Normal distribution with multiplicative standard deviation as $\sigma=-log(\eta_{\rm median})/3$, see Figure~\ref{fig_eta}. Let us remind that in a Log-Normal distribution, the expected value is higher than the true value (i.e., median value) by a factor of $e^{\sigma^2/2}$.
We consider this skewness and recalibrate the inferred expected value of $1/\eta$ to the median value, in order to assign a non-biased $1/\eta$ to the calculation.

\subsection{Luminosity Distance as Redshift Estimator}
\label{sec_dl_z}
In the previous section, we took \dl\ as the redshift estimator to derive the redshift and hence the selection factor $\eta$. Let us denote such inferred redshift as $z_{inf}$. The way to derive the $z_{inf}$ is to take the observed luminosity distance, i.e.,  \dl$(z)$, and find the inverse solution of the integral function based on a fixed cosmological model.

Once the cosmological model is assumed, indirect inference of $z_s$ offers an opportunity to model the BHMF slope as a function of redshift. Therefore, we are able to investigate the second scenario described by the Equation~(\ref{equ_alphaz}) as:
 \begin{equation} \label{equ_lik_alphaz}
 \begin{split}
 P&(\alpha_0, \alpha_1, M_{max}, M_{min}|m_{1},d_L(z)) \propto \\
  &\prod_{i=1}^{total} \hat{P}(m_{1i}, z_{inf,i} |\alpha_0, \alpha_1, M_{max}, M_{min})^{1/\eta}.
  \end{split}
 \end{equation}

 We present our inference for the BHMF using the mock data in the next section.

\vspace{1cm}
\section{Result}\label{sec_result}
We fit the mock data to the BHMF model to infer the distribution of the best-fit parameters. To avoid the bias and estimate the scatter, we adopt the realization approach. In each realization, we simulate a thousand of BBH inspiral GW events and infer the best-fit parameters using minimization of the chi-square objective function. We keep increasing the volume of realizations until the inferred best-fit parameters converged.

\begin{figure*}
\centering
\begin{tabular}{c c c}
\subfloat[assuming $\alpha=0.8$, $M_{min}=5M_{\odot}$ and $M_{max}=50M_{\odot}$.]
{\includegraphics[width=0.3\linewidth]{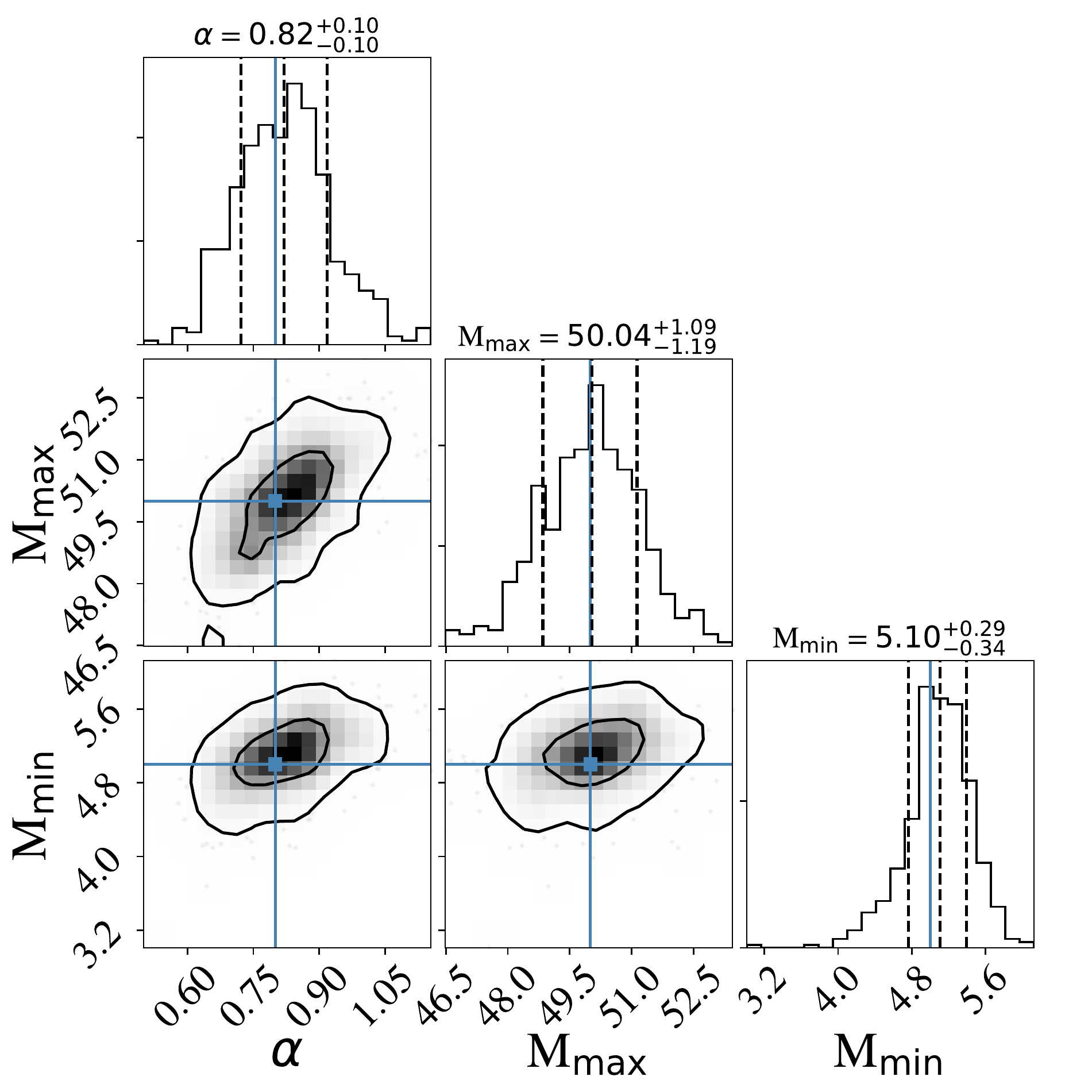}}&
\subfloat[assuming $\alpha=1.6$, $M_{min}=5M_{\odot}$ and $M_{max}=50M_{\odot}$.]
{\includegraphics[width=0.3\linewidth]{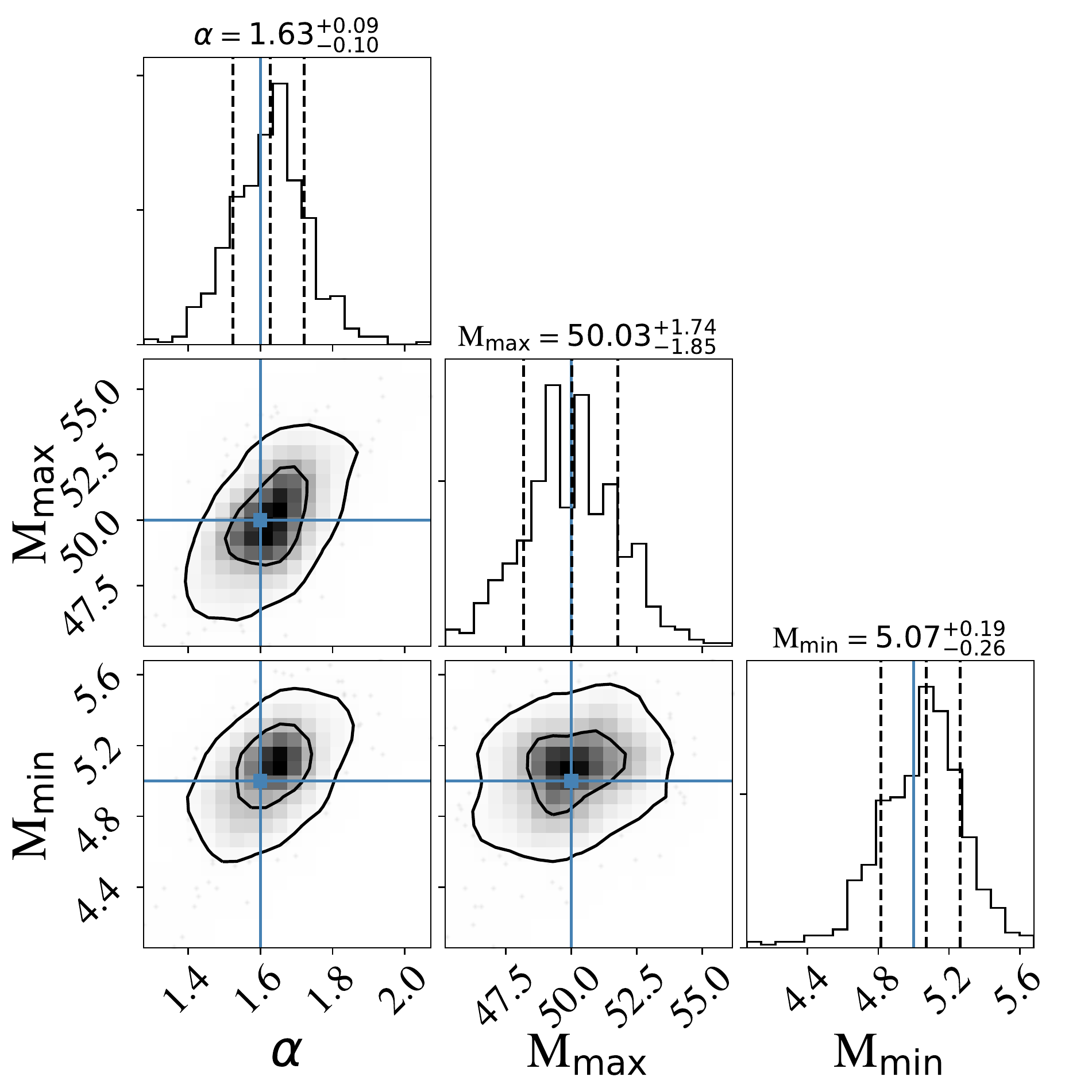}}&
\subfloat[assuming $\alpha=2.4$, $M_{min}=5M_{\odot}$ and $M_{max}=50M_{\odot}$.]
{\includegraphics[width=0.3\linewidth]{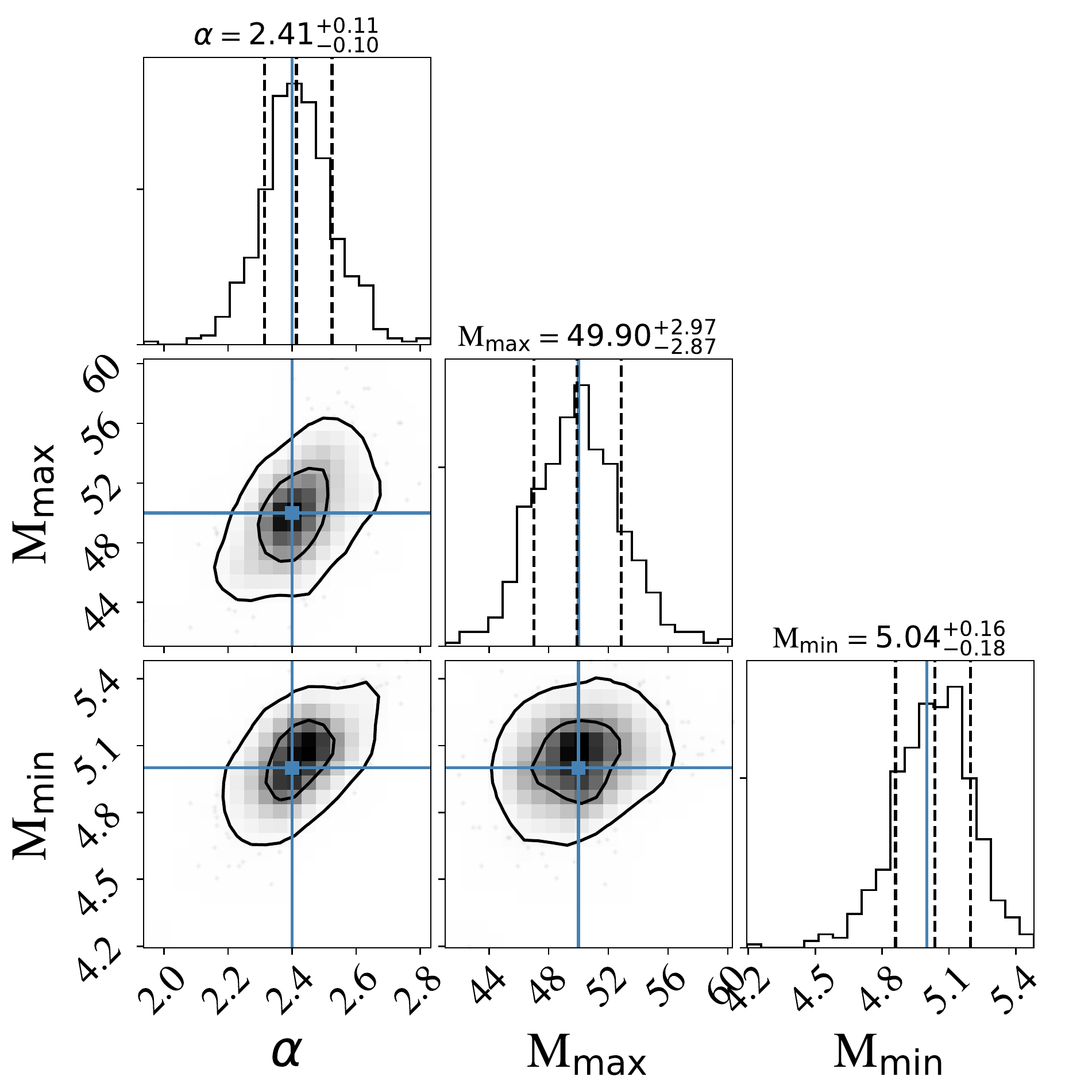}}
\end{tabular}
\caption{
One- and two-dimensional distributions for the best-fitted parameters in the first scenario, based on three sets of parameters with a thousand of BBH inspiral GW events. The BHMF is assumed as a power-law with hard cut at the $M_{min}$ and $M_{max}$, with a constant slope ($\alpha$) across all the redshifts. The blue lines indicate the true value as assumed in the simulation.
}
\label{fig_result_a}
\end{figure*}

In the first scenario, we consider the slope $\alpha$ as a constant. We performed numerical tests assuming three different sets of parameters taking $\alpha$ as 0.8, 1.6 and 2.4, with $M_{min}~=~5M_{\odot}$, $M_{max}~=~50M_{\odot}$. We calculate the likelihood by Equation~(\ref{equ_lik_sf}) to infer the best-fit parameters in each realization. It has been discussed that no black holes with mass over $50M_{\odot}$ are expected from stellar evolution and through supernovae~\citep{Woosley2017, Wiktorowicz2019}.
In Figure~\ref{fig_result_a}, we present the posterior distribution of the inferred parameters for the three parameter sets. We find that all the parameters are recovered accurately which confirms the validity of our method. The uncertainties for the inferred parameters of 68\% confidence interval are $\Delta\alpha\sim0.1$, $\Delta M_{max}\sim1-3M_{\odot}$ and $\Delta M_{min}\sim0.2-0.3M_{\odot}$. We also note that the uncertainty for $\Delta M_{max}$ increases with the increasing of $\alpha$. This is reasonable given that for higher $\alpha$ value, the black mass (i.e., $m_1$) trends to be distributed lower, resulting in a lower constraint power on the high mass end. Clearly, the uncertainty levels cannot be treated as the universal range for the general case, but only apply when the set of the initial parameters is close to the tested ones. Moreover, these uncertainty levels are related to the assumed measurement uncertainties, including ${\cal M}_0$, \dl, and \mone, as discussed in Section~\ref{sec_noiselevel}.

\begin{figure*}
\centering
\begin{tabular}{c c}
\subfloat[assuming $\alpha_0=0.8$, $\alpha_1=0.7$, $M_{min}=5M_{\odot}$ and $M_{max}=50M_{\odot}$.]
{\includegraphics[width=0.4\linewidth]{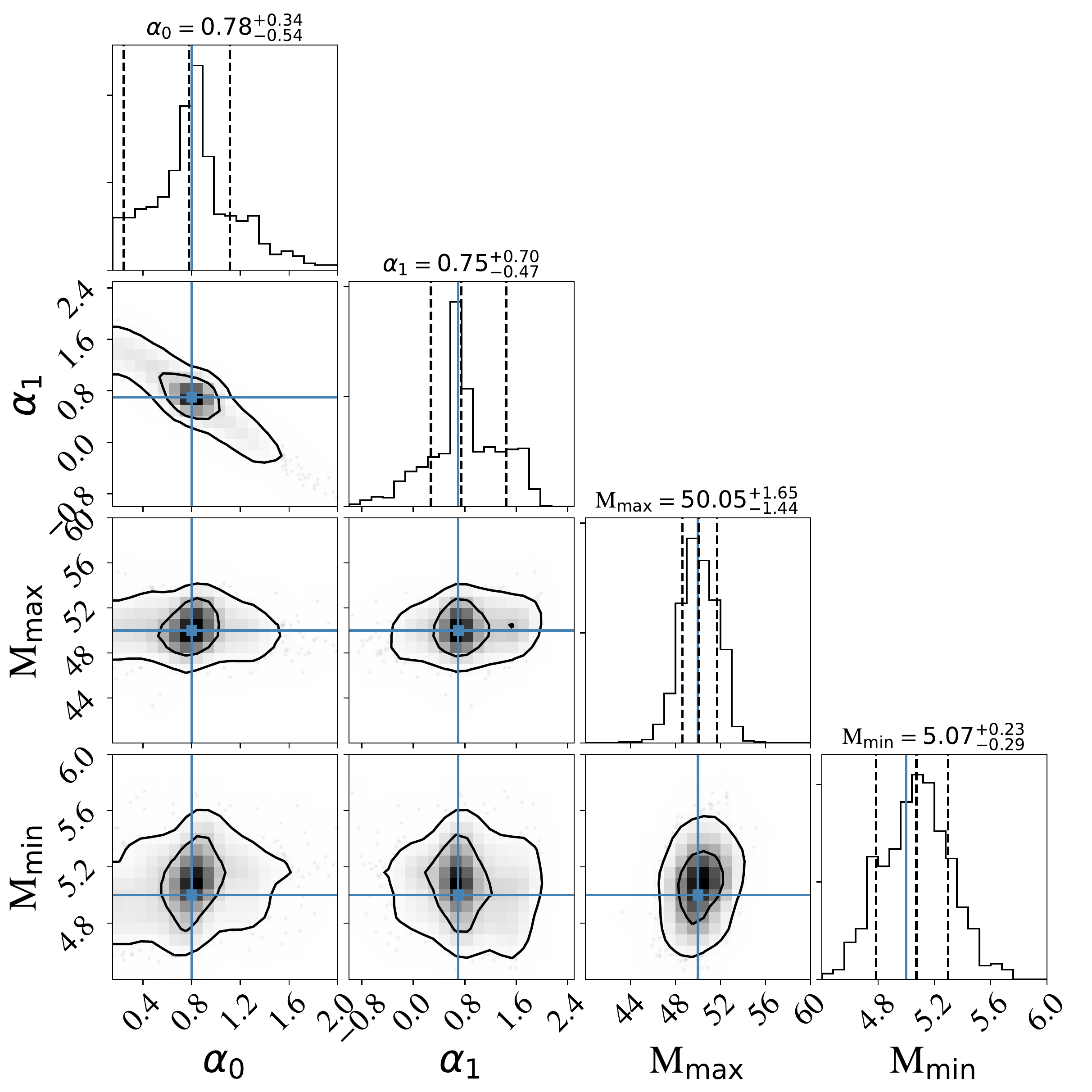}}&
\subfloat[assuming $\alpha_0=1.6$, $\alpha_1=0.7$, $M_{min}=5M_{\odot}$ and $M_{max}=50M_{\odot}$.]
{\includegraphics[width=0.4\linewidth]{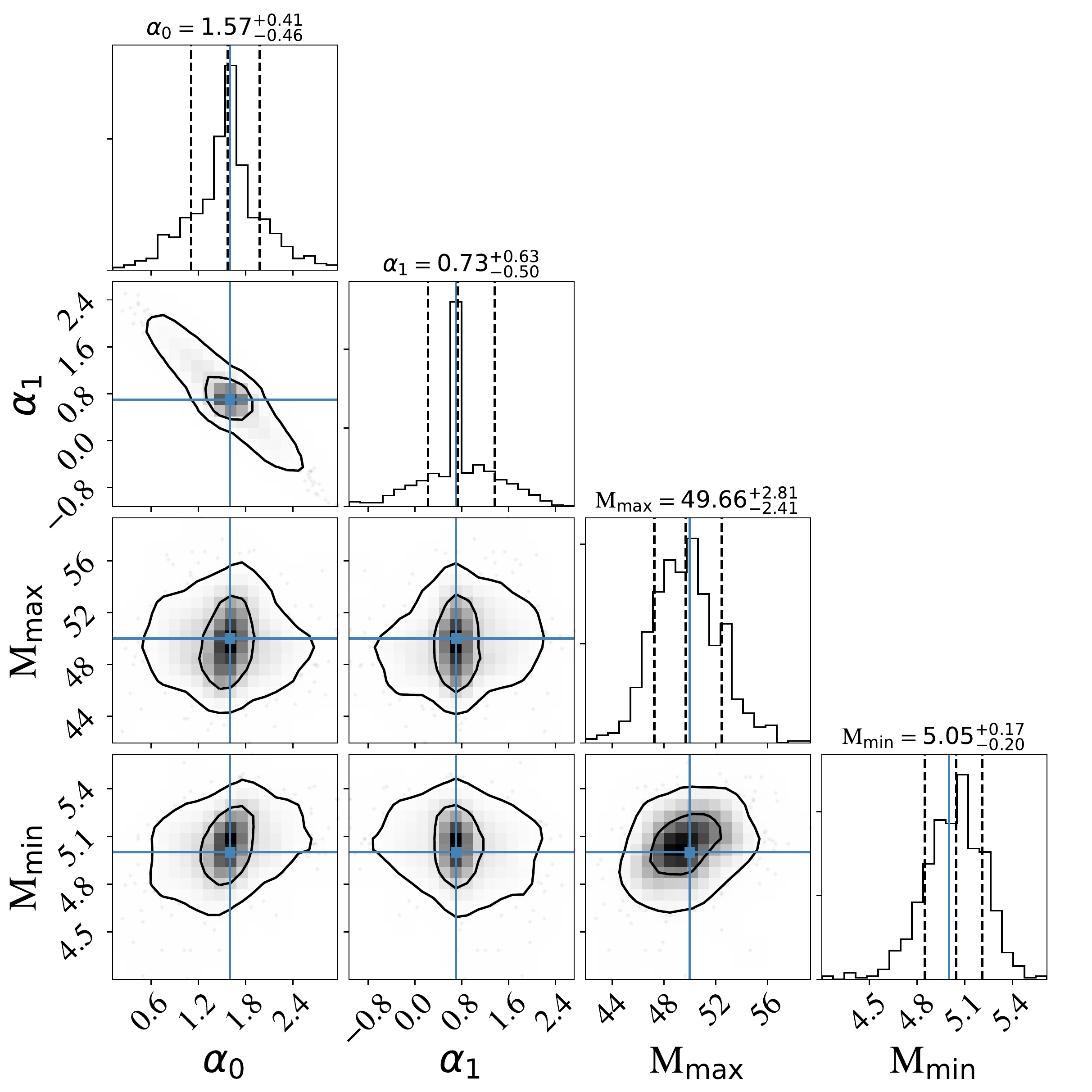}}\\
\subfloat[assuming $\alpha_0=2.4$, $\alpha_1=0.7$, $M_{min}=5M_{\odot}$ and $M_{max}=50M_{\odot}$.]
{\includegraphics[width=0.4\linewidth]{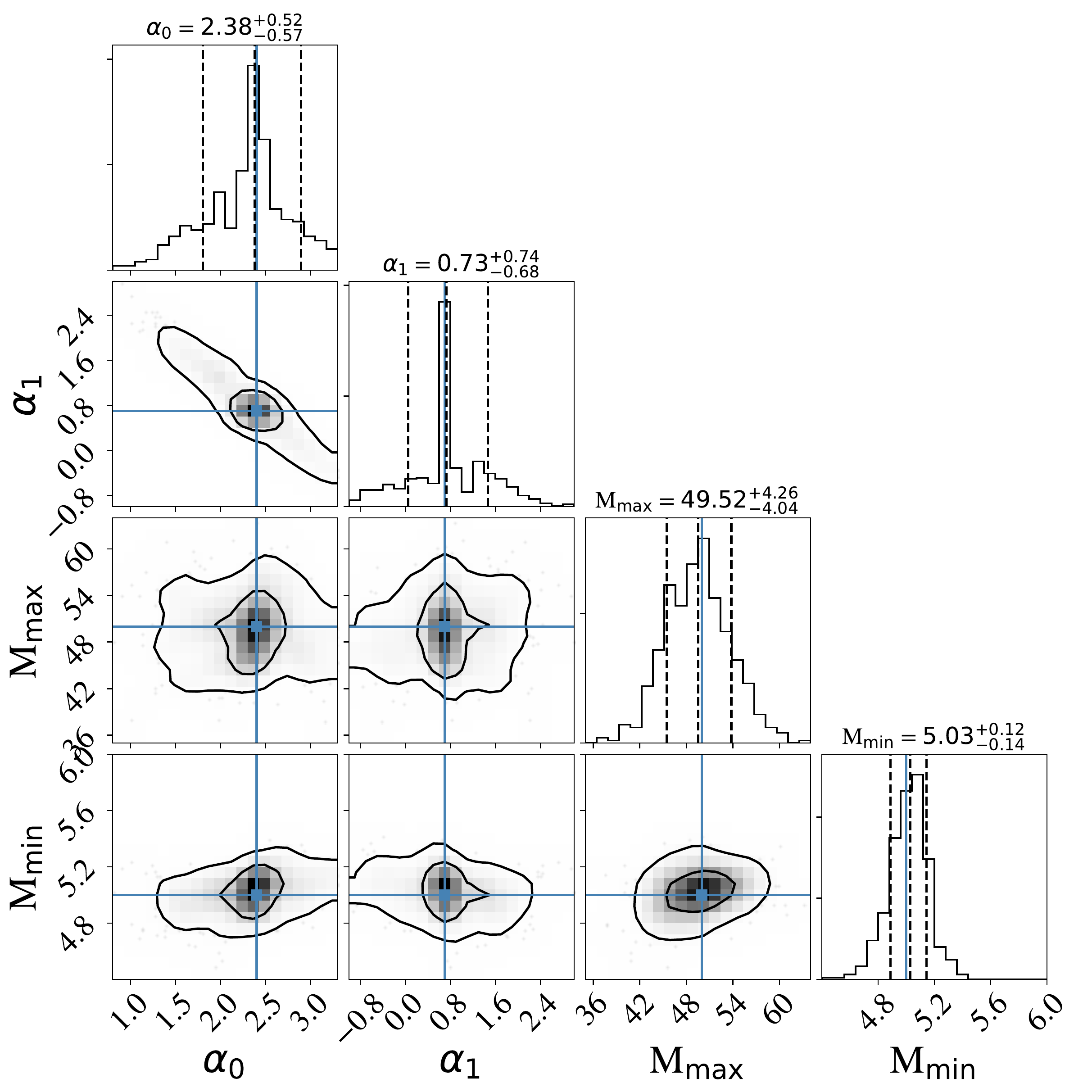}}&
\subfloat[assuming $\alpha_0=1.6$, $\alpha_1=1.2$, $M_{min}=5M_{\odot}$ and $M_{max}=50M_{\odot}$.]
{\includegraphics[width=0.4\linewidth]{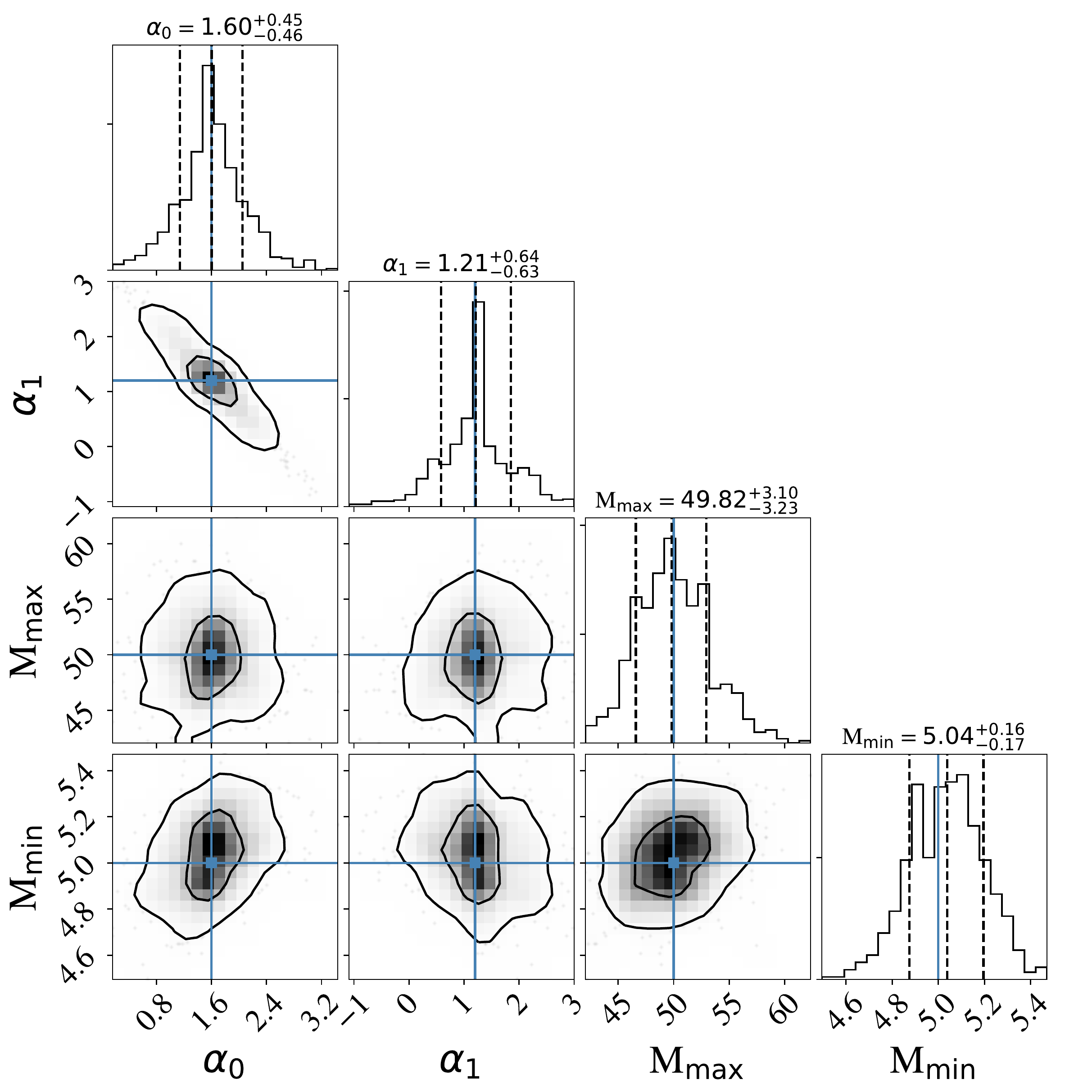}}
\end{tabular}
\caption{
Same as Figure~\ref{fig_result_a} but for the second scenario, where the $\alpha$ of BHMF is evolving with redshift as $\alpha(z) = \alpha_0 + \alpha_1\frac{z}{1+z}$, four sets of parameters assumed.
}
\label{fig_result_b}
\end{figure*}

In the second scenario, the $\alpha$ evolves with redshift according to the Equation~(\ref{equ_alphaz}). We consider four sets of parameters assuming $\alpha_0$ as  0.8, 1.6, 2.4 and $\alpha_1$ including 0.7 and 1.2. We present the results in Figure~\ref{fig_result_b}. One can see that all the assumed parameters could be recovered accurately. With one more parameter included in the second scenario, the uncertainty level are the following: $\Delta\alpha_0\sim0.4$, $\Delta\alpha_1\sim0.5-0.7$, $\Delta M_{max}\sim2-4M_{\odot}$ and $\Delta M_{min}\sim0.2-0.3M_{\odot}$. We note that there is a degeneracy between the $\alpha_0$ and $\alpha_1$, which is understandable given that they are strongly related by the Equation~\ref{equ_alphaz}. However, for the four sets of parameters we tested, this degeneracy does not affect the inferred uncertainty level for $\alpha_0$ and $\alpha_1$.

We highlight that in this second scenario, it is the inferred uncertainty of $\alpha_1$ that matters the most. Our result show that, with only one thousand of GW measurements in the future, the inferred value of $\alpha_1$ would reach to precision of $\Delta\alpha_1\sim0.5-0.7$. Limited by the computing power, we couldn't use numerical test to get a universal uncertainty level for the general case. However, given the four sets of tests as shown the Figure~\ref{fig_result_b}, it is likely to be true that one thousand of GW measurement could distinguish the evolution of BHMF at 1-$\sigma$ confidence level when $\alpha_1$ deviated from 0 by a value of 0.5.
Moreover, we conjecture that the precision of inference is increasing with the sample size as a function of $\sqrt{N}$. Thus, for the four sets of tests, the one year measurements of ET ($\sim10^5$ in total) would decrease the uncertainty levels by a factor of 10.
We also note that the distribution of the best-fitted parameters ($\alpha_0$, $\alpha_1$) does not follow the Gaussian distribution, but rather a large fraction of it is concentrated at the center.

\vspace{1cm}
\section{Conclusion \& Discussion} \label{sec_summary}
The third-generation gravitational wave detector, the Einstein Telescope, is very powerful and capable of detecting $\sim10^5$ GW events per year, with redshift up to $z\sim17$. In this study, we investigated how the detections of the BBH mergers could improve our knowledge of the black hole mass function (BHMF).

We performed the Monte Carlo simulation to estimate the uncertainty level of BHMF parameters inferred from GW signals by BBHs that would be detected by ET. As a starting point, we assumed that the BHMF for the primary BH mass followed a power-law distribution with hard cuts as described by Equation~(\ref{equ_powlaw}). Based on the BBH intrinsic merger rate predicted by {\tt StarTrack}, we randomly simulated the key parameters of the BBH systems, including the chirp masses, redshifts and orientation factors and calculated  their corresponding signal-to-noise ratio \snr\ for the ET. We collected the events whose \snr\ exceeds the detecting threshold and injected Log-Normal noise to the detected parameter, including BH mass, chirp mass, luminosity distance as mock data.

We built up a theoretical framework and explore to use the mock measurements to infer the BHMF. We took into account the measurement uncertainties and the selection effect which would bias the inference. We performed the test using realizations, one thousand GW events adopted per realization, and estimated the distribution of the best-fitted parameters of the BHMF, including the power-law slope $\alpha$, the maximum BH mass $M_{max}$ and the minimum BH mass $M_{min}$ in the first scenario. Furthermore, in the second scenario, we considered  $\alpha$ evolving as a function of redshift as described by Equation~(\ref{equ_alphaz}), and used the luminosity distance as redshift estimator to test this evolution. We summarize our main results as follows:
\begin{enumerate}
\item Using our method based on Equation~(\ref{equ_lik_sf}), the tested parameters are all recovered accurately, as shown in Figure~\ref{fig_result_a}, which confirms the validity of our tests and highlights the importance of correctly considering the measurement uncertainty and selection effect.
\item We assumed $\alpha$ within a scenario in which it is evolving with redshift as $\alpha(z) = \alpha_0 + \alpha_1\frac{z}{1+z}$. Taking the measured \dl\ as redshift estimator and testing with four parameter sets, we are able to recover the true value of $\alpha_1$ accurately, as shown in Figure~\ref{fig_result_b}.
\item Given the fixed sets of parameters, our results show that a volume of one thousand measurements of BBHs events could infer the parameters with uncertainties level at $\Delta\alpha\sim0.1$, $\Delta M_{max}\sim1-2M_{\odot}$ and $\Delta M_{min}\sim0.2-0.3M_{\odot}$ for the first scenario. For the second scenario, the inferred uncertainties are $\Delta\alpha_0\sim0.4$, $\Delta\alpha_1\sim0.5-0.7$, $\Delta M_{max}\sim2-4M_{\odot}$ and $\Delta M_{min}\sim0.2-0.3M_{\odot}$. In the future, the one year detection rate of ET ($\sim10^5$ in total) would increase the sample size by a factor of 100. According to the fact that the precision of the inference increases with the sample size, as a function of $\sqrt{N}$, we conclude that one year BBH sample by ET would be able to deliver the parameters with uncertainties reduced by a factor of 10 with respect to these reported in this paper.
\end{enumerate}

We point out a few circumstances, which might weaken generality this work. First, we have adopted a template of intrinsic BBH merger rate based on the predictions by a standard model in {\tt StarTrack}, which can be different from the realistic one. Of course, the intrinsic BBH merger rate is unknown yet, which is related to lack of detailed knowledge of different elements such as BBH masses, explosion mechanism, the metallicity history and the time delay distribution. With a different template of BBH merger rate, the simulated mock events (i.e., the ones in the Table~\ref{tab_GW_mock_data}) would follow a different redshift distribution, which could slightly change the inference of the uncertainties of the inferred parameters. Second, for the sake of simplicity, we simulated the value of the secondary BH mass \mtwo\ by assuming that two masses of BBHs have independent distributions, which probably is not exactly true. One can expect that these two limitations would affect the prediction of the yearly detection rate of the GW events and their redshift distribution; however, their influence on our final inferred contours of BHMF (i.e., Figure~\ref{fig_result_a} and~\ref{fig_result_b}) is likely not very significant.
At last, the numerical tests done in this work confirmed the validity of our method. However, limited by the sets of tests, the uncertainty of inferred parameters  in both scenarios applies directly to the fixed sets of parameters and shouldn't be applied to the general case.

In this work, we focused on the inference of the BHMF using the mass properties by the BBH. However, it is worth to note that our approach could be extended to address other problems. For example, one could infer the spin of BH \citep{Abbott2018b}, the mass function for the binary of NS-NS, NS-BH system, though these events are detectable at lower redshift ($z<4$). In addition, using the luminosity as redshift estimator, one should also be able to reconstruct the BBH intrinsic merger rate \citep{Fishbach2018}, and the cosmological parameters.

\acknowledgments
We thank Hosek Jr., M.W for the useful discussion.

This work was supported by the National Natural Science Foundation of China under grant Nos. 11633001 and 11920101003, the Strategic Priority Research Program of the Chinese Academy of Sciences, grant No. XDB23000000, and the Interdiscipline Research Funds of Beijing Normal University.

X. Ding acknowledges support by China Postdoctoral Science Foundation Funded Project (No. 2017M622501).
M.B. was supported by the Key Foreign Expert Program for the Central Universities No. X2018002.
K. Liao was supported by the National Natural Science Foundation of China (NSFC) No. 11973034.
\software{  {\sc corner}, \citep{corner},
        Matplotlib \citep{Matplotlib},
        and standard Python libraries.
        }




\end{document}